\documentclass[]{article}
\usepackage{amsmath}
\usepackage{amssymb}
\usepackage{amsthm}
\usepackage{amsfonts}
\usepackage{pstricks}

\newtheorem{theorem}{Theorem}[section]
\newtheorem{lemma}[theorem]{Lemma}

\newtheorem{definition}[theorem]{Definition}

\newcommand{\GL}[1]{\ensuremath{\mathrm{GL}\left(#1\right)}}
\newcommand{\Sym}[1]{\ensuremath{\mathrm{Sym}\left(#1\right)}}
\newcommand{\Aut}[1]{\ensuremath{\mathrm{Aut}\left(#1\right)}}
\newcommand{\Vertex}[1]{\ensuremath{V\left(#1\right)}}
\newcommand{\Edge}[1]{\ensuremath{E\left(#1\right)}}
\newcommand{\pcyc}{\mathbb{Z}/p\mathbb{Z}}
\begin{document}
\title{Representing groups on graphs}
\author{Sagarmoy Dutta and Piyush P Kurur \\
Department of Computer Science and Engineering,\\
Indian Institute of Technology Kanpur,\\
Kanpur, Uttar Pradesh, India 208016\\
\tt{sagarmoy@cse.iitk.ac.in, ppk@cse.iitk.ac.in}}

\date{}
\maketitle

\begin{abstract}
  In this paper we formulate and study the problem of representing
  groups on graphs. We show that with respect to polynomial time
  turing reducibility, both abelian and solvable group
  representability are all equivalent to graph isomorphism, even when
  the group is presented as a permutation group via generators. On the
  other hand, the representability problem for general groups on trees
  is equivalent to checking, given a group $G$ and $n$, whether a
  nontrivial homomorphism from $G$ to $S_n$ exists.  There does not
  seem to be a polynomial time algorithm for this problem, in spite of
  the fact that tree isomorphism has polynomial time algorithm.
\end{abstract}

\section{Introduction}

Representation theory of groups is a vast and successful branch of
mathematics with applications ranging from fundamental physics to
computer graphics and coding theory \cite{repRealworld}. Recently
representation theory has seen quite a few applications in computer
science as well.  In this article, we study some of the questions
related to representation of finite groups on graphs.

A representation of a group $G$ usually means a linear representation,
i.e. a homomorphism from the group $G$ to the group $\GL{V}$ of
invertible linear transformations on a vector space $V$.  Notice that
$\GL{V}$ is the set of \emph{symmetries} or \emph{automorphisms} of
the vector space $V$. In general, by a representation of $G$ on an
object $X$, we mean a homomorphism from $G$ to the automorphism group
of $X$.  In this article, we study some computational problems that
arise in the representation of \emph{finite groups} on graphs. Our
interest is the following group representability problem: Given a
group $G$ and a graph $X$, decide whether $G$ has a nontrivial
representation on $X$. As expected this problem is closely connected
to graph isomorphism: We show, for example, that the graph isomorphism
problem reduces to representability of abelian groups. In the other
direction we show that even for solvable groups the representability
on graphs is decidable using a graph isomorphism oracle. Surprisingly
the non-solvable version of this problem seems to be harder than graph
isomorphism. For example, we were able to show that representability of
groups on trees, a class of graphs for which isomorphism is decidable
in polynomial time, is as hard as checking whether, given an integer
$n$ and a group $G$, the symmetric group $S_n$ has a nontrivial
subgroup homomorphic to $G$, a problem for which no polynomial time
algorithm is known.

\section{Background}

In this section we review the group theory required for the rest of
the article. Any standard text book on group theory, for example the
one by Hall~\cite{hall}, will contain the required results.

We use the following standard notation: The identity of a group $G$ is
denoted by $1$. In addition $1$ also stands for the singleton group
consisting of only the identity. For groups $G$ and $H$, $H \leq G$
(or $G\geq H$) means that $H$ is a subgroup of $G$. Similarly by $H
\unlhd G$ (or $G\unrhd H$) we mean $H$ is a \emph{normal subgroup} of
$G$.

Let $G$ be any group and let $x$ and $y$ be any two elements. By the
\emph{commutator} of $x$ and $y$, denoted by $[x,y]$, we mean
$xyx^{-1}y^{-1}$. The \emph{commutator subgroup} of $G$ is the group
generated by the set $\{[x,y]|x,y \in G\}$. We denote the commutator
subgroup of $G$ by $G'$. The following is a well known result in group
theory~\cite[Theorem 9.2.1]{hall}

\begin{theorem}
The commutator subgroup $G'$ is a normal subgroup of $G$ and $G/G'$ is
abelian. Further for any normal subgroup $N$ of $G$ such that $G/N$ is
abelian, $N$ contains $G'$ as a subgroup.
\end{theorem}

A group is \emph{abelian} if it is commutative, i.e. $g h = hg$ for
all group elements $g$ and $h$. A group $G$ is said to be
\emph{solvable}~\cite[Page 138]{hall} if there exists a decreasing
chain of groups $G = G_0 \rhd G_1 \ldots \rhd G_t = 1$ such that
$G_{i+1}$ is the commutator subgroup of $G_i$ for all $0 \leq i < t$.

An important class of groups that play a crucial role in graph
isomorphism and related problems are permutation groups. We follow the
notation of Wielandt~\cite{wielandt64finite} for permutation groups.
Let $\Omega$ be a finite set. The \emph{symmetric group} on $\Omega$,
denoted by $\Sym{\Omega}$, is the group of all permutations on the set
$\Omega$.  By a \emph{permutation group} on $\Omega$ we mean a
subgroup of the symmetric group $\Sym{\Omega}$. For any positive
integer $n$, we will use $S_n$ to denote the symmetric group on
$\{1,\ldots, n\}$. Let $g$ be a permutation on $\Omega$ and let
$\alpha$ be an element of $\Omega$. The image of $\alpha$ under $g$
will be denoted by $\alpha^g$. For a permutation group $G$ on
$\Omega$, the orbit of $\alpha$ is denoted by $\alpha^G$. Similarly if
$\Delta$ be a subset of $\Omega$ then $\Delta^g$ denotes the set $\{
\alpha^g | \alpha \in \Delta \}$.

Any permutation group $G$ on $n$ symbols has a generating set of size
at most $n$. Thus for computational tasks involving permutation groups
it is assumed that the group is presented to the algorithm via a small
generating set. As a result, by efficient algorithms for permutation
groups on $n$ symbols we mean algorithms that take time polynomial in
the size of the generating set and $n$.

Let $G$ be a subgroup of $S_n$ and let $G^{(i)}$ denote the subgroup
of $G$ that fixes pointwise $j \leq i$, i.e. $G^{(i)} = \{ g | j^g =
j, 1 \leq j \leq i \}$. Let $C_i$ denote a right \emph{transversal},
i.e.  the set of right coset representative, for $G^{(i)}$ in
$G^{(i-1)}$. The $\cup_i C_i$ is a generating set for $G$ and is
called the \emph{strong generating set} for $G$. The corner stone for
most polynomial time algorithms for permutation group is the
Schreier-Sims~\cite{sims70computational,sims78some,furst80polynomialtime}
algorithm for computing the \emph{strong generating set} of a
permutation group $G$ given an arbitrary generating set. Once the
strong generating set is computed, many natural problems for
permutation groups can be solved efficiently. We give a list of them
in the next theorem.

\begin{theorem}\label{thm-perm-polytime}
  Given a generating set for $G$ there are polynomial time algorithms
  for the following task.
  \begin{enumerate}
  \item Computing the strong generating set.
  \item Computing the order of $G$.
  \end{enumerate}
\end{theorem}

By a graph we mean a \emph{finite undirected graph}. For a graph $X$,
$\Vertex{X}$ and $\Edge{X}$ denotes the set of vertices and edges
respectively and $\Aut{X}$ denotes the group of all
automorphisms of $X$, i.e. permutations on $\Vertex{X}$ that maps
edges to edges and non-edges to non-edges. 

\begin{definition}[Representation]
  A representation $\rho$ from a group $G$ to a graph $X$ is a
  homomorphism from $G$ to the automorphism group $\Aut{X}$ of $X$.
\end{definition}

Alternatively we say that $G$ acts on (the right) of $X$ via the
representation $\rho$. When $\rho$ is understood, we use $u^g$ to
denoted $u^{\rho(g)}$. 

A representation $\rho$ is \emph{trivial} if all the elements of $G$
are mapped to the identity permutation. A representation $\rho$ is
said to be \emph{faithful} if it is an injection as well. Under a
faithful action $G$ can be thought of as a subgroup of the
automorphism group. We say that $G$ is \emph{representable} on $X$
 if there is a nontrivial representation from $G$ to
$X$. We now define the following natural computational problem.

\begin{definition}[Group representability problem]
  Given a group $G$ and a graph $X$ decide whether $G$ is
  representable on $X$ nontrivially.
\end{definition}

We will look at various restrictions of the above problem. For example,
we study the abelian (solvable) group representability problem where
our input groups are abelian (solvable). We also study the group
representability problem on trees, by which we mean group
representability where the input graph is a tree.

Depending on how the group is presented to the algorithm, the
complexity of the problem changes. One possible way to present $G$ is
to present it as a permutation group on $m$ symbols via a generating
set. In this case the input size is $m + \# V(X)$. On the other hand,
we can make the task of the algorithm easier by presenting the group
via a multiplication table. In this paper we mostly assume that the
group is in fact presented via its multiplication table. Thus
polynomial time means polynomial in $\# G$ and $\# V(X)$. However for
solvable representability problem, our results extend to the case when
$G$ is a permutation group presented via a set of generators.

We now look at the following closely related problem that occurs
when we study the representability of groups on trees.

\begin{definition}[Permutation representability problem] 
  \label{def-perm-representability}
  Given a group $G$ and an integer $n$ in unary, check whether there is
  a homomorphism from $G$ to $S_n$.
\end{definition}

\subsection*{Overview of the results}

Our first result is to show that graph isomorphism reduces to abelian
representability problem. In fact we show that graph isomorphism
reduces to the representability of prime order cyclic groups on
graphs. Next we show that solvable group representability problem
reduces to graph isomorphism problem. Thus as far as polynomial time
Turing reducibility is concerned abelian group representability and
solvable group representability are all equivalent to graph
isomorphism. As a corollary we have, solvable group representability on
say bounded degree graphs or bounded genus graphs are all in
polynomial time.

We then show that group representability on trees is equivalent to
permutation representability
(Definition~\ref{def-perm-representability}). This is in contrast to
the corresponding isomorphism problem because for trees, isomorphism
testing is in polynomial time whereas permutation representability
problem does not appear to have a polynomial time algorithm.

\section{Abelian representability}

In this section we prove that the graph isomorphism problem reduces to
abelian group representability on graph. Given input graphs $X$ and
$Y$ of $n$ vertices each and any prime $p > n$, we construct a graph
$Z$ of exactly $p \cdot n$ vertices such that $X$ and $Y$ are
isomorphic if and only if the cyclic group of order $p$ is
representable on $Z$. Since for any integer $n$ there is a
prime $p$ between $n$ and $2n$ (Bertrand's conjecture), the above
constructions gives us a reduction from the graph isomorphism problem
to abelian group representability problem.

For the rest of the section, fix the input graphs $X$ and $Y$. Our
task is to decide whether $X$ and $Y$ are isomorphic.  Firstly we
assume, without loss of generality, that the graphs $X$ and $Y$ are
connected, for otherwise we can take their complement graphs $X'$ and
$Y'$, which are connected and are isomorphic if and only if $X$ and
$Y$ are isomorphic. Let $n$ be the number of vertices in $X$ and $Y$
and let $p$ be any prime greater than $n$. Consider the graph $Z$
which is the disjoint union of $p$ connected components
$Z_1,\ldots,Z_p$ where, for each $1 \leq i < p$, each $Z_i$ is an
isomorphic copy of $X$ and $Z_p$ is an isomorphic copy of $Y$. First
we prove the following lemma.

\begin{lemma}\label{lem-gigrepf}
  If $X$ and $Y$ are isomorphic then $\pcyc$ is representable
  on $Z$.
\end{lemma}
\begin{proof}
  Clearly it is sufficient to show that there is an order $p$
  automorphism for $Z$. Let $h$ be an isomorphism from $X$ to $Y$. For
  every vertex $v$ in $X$, let $v_i$ denote its copy in
  $Z_i$. Consider the bijection $g$ from $V(Z)$ to itself defined as
  follows: For all vertices $v$ in $V(X)$ and each $1 \leq i < p - 2$,
  let $v_i^g = v_{i+1}$. Further let $g$ map $v_{p-1}$ to $v^h$ and
  $v^h$ to $v_1$. It is easy to verify that $g$ is an automorphism of
  $Z$ and has order $p$.
\end{proof}

We now prove the converse
\begin{lemma}\label{lem-gigrepb}
  If $\pcyc$ can be represented on $Z$ then $X$ and $Y$ are
  isomorphic.
\end{lemma}
\begin{proof}
  If $\pcyc$ can be represented on $Z$ then there exists a nonidentity
  automorphism $g$ of $Z$ such that order of $g$ is $p$. We consider
  the action of the cyclic group $H$, generated by $g$, on $V(X)$. Since
  $g$ is nontrivial, there exists at least one $H$-orbit $\Delta$ of
  $V(X)$ which is of cardinality greater than $1$. However by orbit
  stabiliser formula \cite[Theorem 3.2]{wielandt64finite}, $\# \Delta$
  divides $\# H = p$. Since $p$ is prime, $\Delta$ should be of
  cardinality $p$.

  We prove that no two vertices of $\Delta$ belong to the same
  connected component. Assume the contrary and let $\alpha$ and
  $\beta$ be two elements of $\Delta$ which also belong to the same
  connected component of $Z$. There is some $0 < t < p$ such that
  $\alpha^{g^t} = \beta$. We assume further, without loss of
  generality, that $t=1$, for otherwise we replace $g$ by the
  automorphism $g^t$, which is also of order $p$, and carry out the
  argument. Therefore $\alpha^g = \beta$ lie in the same component of
  $Z$. It follows then that, for each $0 \leq i \leq p -1 $, the element
  $\alpha_i = \alpha^{g^i}$ is in the same component of $Z$, as
  automorphisms preserve edges and hence paths. However this means
  that there is a component of $Z$ that is of cardinality at least
  $p$. This is a contradiction as each component of $Z$ has at most $n
  < p$ vertices as they are copies of either $X$ or $Y$.

  It follows that there is some $1 \leq i < p$, for which $g$ must map
  at least one vertex of the component $Z_i$ to some vertex of
  $Z_p$. As a result the automorphism $g$ maps the entire component
  $Z_i$ to $Z_p$.  Therefore the components $Z_i$ and $Z_p$ are
  isomorphic and so are their isomorphic copies $X$ and $Y$.
\end{proof}

Given two graphs $X$ and $Y$ of $n$ vertices we find a prime $p$ such
that $n < p < 2n$, construct the graph $Z$ and construct the
multiplication table for $\pcyc$. This requires only logarithmic space
in $n$. Using Lemmas~\ref{lem-gigrepf} and \ref{lem-gigrepb} we have
the desired reduction.

\begin{theorem}
  The graph isomorphism problem logspace many-one reduces to abelian
  group representability problem.
\end{theorem}

\section{Solvable representability problem}

In the previous section we proved that abelian group representability
is at least as hard as graph isomorphism. In this section we show that
solvable group representability is polynomial time Turing reducible to
the graph isomorphism problem. We claim that a solvable group $G$ is
representable on $X$ if and only if $\# \Aut{X}$ and $\# G/G'$ have a
common prime factor, where $G'$ is commutator subgroup of $G$. We do
this in two stages.

\begin{lemma}\label{lem-solf}
  A solvable group $G$ can be represented on a graph $X$ if $\#G/G'$
  and $\#\Aut{X}$ have a common prime factor.
\end{lemma}
\begin{proof}

Firstly notice that it suffices to prove that there is a nontrivial
homomorphism, say $\rho$, from $G/G'$ to $\Aut{X}$. A nontrivial
representation for $G$ can be obtained by composing the natural
quotient homomorphism from $G$ \emph{onto} $G/G'$ with $\rho$.

Recall that the quotient group $G/G'$ is an abelian group and hence
can be represented on $X$ if for some prime $p$ that divides $\#
G/G'$, there is an order $p$ automorphism for $X$. However by the
assumption of the theorem, there is a common prime factor, say $p$, of
$\# G/G'$ and $\# \Aut{X}$. Therefore, by Cayley's theorem there is an
order $p$ element in $\Aut{X}$. As a result, $G/G'$ and hence $G$ is
representable on $X$.
\end{proof}

To prove the converse, for the rest of the section fix the input, the
solvable group $G$ and the graph $X$. Consider any nontrivial
homomorphism $\rho$ from the group $G$ to $\Aut{X}$. Let $H \leq
\Aut{X}$ denote the image of the group $G$ under $\rho$. We will from
now on consider $\rho$ as an automorphism from $G$ \emph{onto}
$H$. Since the subgroup $H$ is the homomorphic image of $G$, $H$
itself is a solvable group.

\begin{lemma}\label{comm}
The homomorphism $\rho$ maps the commutator subgroup $G'$ of $G$
\emph{onto} the commutator subgroup $H'$.
\end{lemma}
\begin{proof}
  First we prove that $\rho(G') \leq H'$. For this notice that for all
  $x$ and $y$ in $G$, since $\rho$ is a homomorphism, $\rho([x,y]) =
  [\rho(x),\rho(y)]$ is an element of $H'$. As $G'$ is generated by
  the set $\{ [x,y] | x,y \in G\}$ of all commutators, $\rho(G') \leq
  H'$.  To prove the converse notice that $\rho$ is a surjection on
  $H$.  Therefore for any element $h$ of $H$, we have element $x_h$ of
  $G$ such that $\rho(x_h) = h$. Consider the commutator $[g,h]$ for
  any two elements $g$ and $h$ of $H$. We have $\rho([x_g,x_h]) =
  [g,h]$.  This proves that all the commutators of $H$ are in the
  image of $G'$ and hence $\rho(G') \geq H'$.
\end{proof}

We have the following result about solvable groups that directly
follows from the definition of solvable groups \cite[Page 138]{hall}.

\begin{lemma}\label{lem-nontrivial-govergprime}
  Let $G$ be any nontrivial solvable group then its commutator
  subgroup $G'$ is a strict subgroup of $G$.
\end{lemma}
\begin{proof}
  By the definition of solvable groups, there exist a chain $G = G_0
  \rhd G_1 \ldots \rhd G_t = 1$ such that $G_{i+1}$ is the commutator
  subgroup of $G_i$ for all $0 \leq i < t$. If $G=G'=G_1$ then $G=G_i$
  for all $0 \leq i \leq t$ implying $G=1$
\end{proof}

We are now ready to prove the converse of Lemma~\ref{lem-solf}.

\begin{lemma}\label{lem-solb}
  Let $G$ be any solvable group and let $X$ be any graph, then $G$ is
  representable on graph $X$ if $\#G/G'$ and $\#\Aut{X}$ have a common
  prime factor.
\end{lemma}
\begin{proof}
  Let $\rho$ be any nontrivial homomorphism from $G$ to $\Aut{X}$, and
  let $H$ be the image of group $G$ under this homomorphism. Since the
  commutator subgroup $G'$ is strictly contained in the group $G$
  (Lemma~\ref{lem-nontrivial-govergprime}), order of the quotient
  group $\#G/G' >1$. Furthermore, the image group $H$ itself is
  solvable and nontrivial, as it is the image of a solvable group $G$
  under a nontrivial homomorphism. Therefore, the commutator subgroup
  $H'$ is strictly contained in $H$ implying $\#H/\#H'>1$.

  Consider the homomorphism $\tilde{\rho}$ from $G$ \emph{onto} $H/H'$
  defined as $\tilde{\rho}(g) = \rho(g) H'$. Since $\rho$ maps $G'$
  onto $H'$, we have that $G'$ is in the kernel of
  $\tilde{\rho}$. Therefore, $\tilde{\rho}$ can be \emph{refined} to a
  map from $G/G'$ \emph{onto} $H/H'$. Clearly the prime factors of $\#
  H/H'$ are all prime factors of $\# G/G'$. However, any prime factor
  of $\# H/H'$ is a prime factor of $\Aut{X}$, as both $H$ and $H'$
  are subgroups of $\Aut{X}$. Therefore, the orders of $G/G'$ and
  $\Aut{X}$ have a common prime factor.
\end{proof}

The order of the automorphism group of the input graph $X$ can be
computed in polynomial time using an oracle to the graph isomorphism
problem~\cite{mathon79note}. Further since the automorphism group is a
subgroup of $S_n$, where $n$ is the cardinality of $V(X)$, all its
prime factors are less than $n$ and hence can be determined. Also
since $G$ is given as a table, its commutator subgroup $G'$ can be
computed in polynomial time and the prime factors of $\# G/G'$ can
also be similarly determined.  Therefore we can easily check, given
the group $G$ via its multiplication table and the graph $X$, whether
the order of the quotient group $G/G'$ has common factors with the
order of $\Aut{X}$. We thus have the following theorem.

\begin{theorem}\label{thm-solvable-to-gi}
  The problem of deciding whether a solvable group can be represented
  on a given graph reduces to graph isomorphism problem.
\end{theorem}

For the reduction in the above theorem to work, it is sufficient to
compute the order of $G$ and its commutator subgroup $G'$. This can be
done even when the group $G$ is presented as a permutation group on
$m$ symbols via a generating set. To compute $\# G$ we can compute the
strong generating set of $G$ and use
Theorem~\ref{thm-perm-polytime}. Further given a generating set for
$G$, a generating set for its commutator subgroup $G'$ can be compute
in polynomial time~\cite[Theorem 4]{furst80polynomialtime}. Therefore,
the order of $G/G'$ can be computed in polynomial time given the
generating set for $G$. Furthermore, $G$ and $G'$ are subgroups of
$S_m$ and hence all their prime factors are less than $m$ and can
be determined. We can then check whether $\# G/G'$ has any common
prime factors with $\# \Aut{X}$ just as before using the graph isomorphism
oracle. Thus we have the following theorem.

\begin{theorem}
  The solvable group representability problem, where the group is
  presented as a permutation group via a generating set, reduces to
  the graph isomorphism problem via polynomial time Turing reduction.
\end{theorem}

\section{Representation on tree}\label{sect-tree-representability}

In this section we study the representation of groups on trees. It is
known that isomorphism of trees can be tested in polynomial
time~\cite{babai83canonical}. However we show that the group
representability problem over trees is equivalent to permutation
representability problem (Definition~\ref{def-perm-representability}),
a problem for which, we believe, there is no polynomial time
algorithm.

\begin{figure}[h!]
\begin{center}
\begin{pspicture}(3,1.3)
\psline{*-*}(1.5,1.2)(0.1,0.1)
\psline{-*}(1.5,1.2)(0.7,0.1)
\psline{-*}(1.5,1.2)(1.3,0.1)
\psline{-*}(1.5,1.2)(2.9,0.1)
\rput(1.8,0.1){$\ldots$}
\end{pspicture}
\end{center}
\caption{Tree with automorphism group $S_n$}\label{fig-tree-with-sn}
\end{figure}
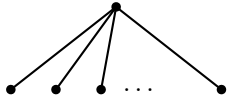

Firstly, to show that permutation representability problem is
reducible to group representability problem on trees, it is sufficient
to construct, given and integer $n$, a tree whose automorphism group
is $S_n$. Clearly a tree with $n$ leaves, all of which is connected to
the root, gives such a tree (see
Figure~\ref{fig-tree-with-sn}). Therefore we have the following lemma.
\begin{lemma}\label{lemsn2tree}
Permutation representability reduces to representability on tree.
\end{lemma}

\begin{figure}[h!]
\begin{center}
\begin{pspicture}(9,2.2)
\psline{*-*}(1,1.4)(3,1.4)
\pspolygon(1,1.4)(0.3,0.1)(1.7,0.1)
\pspolygon(3,1.4)(2.3,0.1)(3.7,0.1)
\rput[b](1,1.5){$\alpha$} \rput[b](3,1.5){$\beta$}
\psline{*-*}(6,1.4)(7,1.8)
\psline{-*}(7,1.8)(8,1.4)
\pspolygon(6,1.4)(5.3,0.1)(6.7,0.1)
\pspolygon(8,1.4)(7.3,0.1)(8.7,0.1)
\rput[b](6,1.5){$\alpha$} \rput[b](8,1.5){$\beta$} \rput[b](7,1.9){$\gamma$}
\end{pspicture}
\end{center}
\caption{Minimal orbit has two elements}\label{fig-maximal-orbit}
\end{figure}
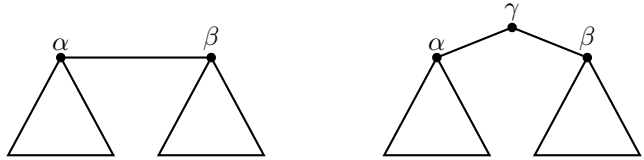

To prove the converse, we first reduce the group representability
problem on an arbitrary tree to the problem of representability on a
rooted tree. We then do a divide and conquer on the structure of the
rooted tree using the permutation representability oracle. The main
idea behind this reduction is Lemma~\ref{lem-edge} where we show that for any
tree $T$, either there is a vertex which is fixed by all automorphism,
in which case we can choose this vertex as the root, or there are two
vertices $\alpha$ and $\beta$ connected by an edge which together
forms an orbit under the action of $\Aut{T}$, in which case we can add
a dummy root (see Figure~\ref{fig-maximal-orbit}) to make it a rooted
tree without changing the automorphism group.

For the rest of the section fix a tree $T$.  Let $\Delta$ be an orbit
in the action of $\Aut{T}$ on $V(T)$.  We define the graph $T_\Delta$
as follows: A vertex $\gamma$ (or edge $e$) of $T$ belongs to
$T_\Delta$ if there are two vertices $\alpha$ and $\beta$ in $\Delta$
such that $\gamma$ (or $e$) is contained in the path from $\alpha$ to
$\beta$. It is easy to see that $T_\Delta$ contains paths between any
two vertices of $\Delta$. Any vertex in $T_{\Delta}$ is connected to
some vertex in $\Delta$ and all vertices in $\Delta$ are connected in
$T_{\Delta}$ which implies $T_{\Delta}$ is connected. Furthermore
$T_\Delta$ has no cycle, as its edge set is a subset of the edge set
of $T$. Therefore $T_\Delta$ is a tree.

\begin{lemma}
  Let $g$ be any automorphism of $T$ and consider any vertex $\gamma$
  (or edge $e$) of $T_\Delta$. Then the vertex $\gamma^g$ (or edge
  $e^g$) is also in $T_\Delta$.
\end{lemma}
\begin{proof}
  Since $\gamma$ (or $e$) is present in $T_\Delta$, there exists
  $\alpha$ and $\beta$ in $\Delta$ such that $\gamma$ (or $e$) is in
  the path between $\alpha$ and $\beta$. Also since automorphisms
  preserve paths, $\gamma^g$ (or $e^g$) is in the path from
  $\alpha^g$ to $\beta^g$.
\end{proof}

\begin{lemma}\label{lemleaf}
  The orbit $\Delta$ is precisely the set of leaves of $T_{\Delta}$.
\end{lemma}
\begin{proof}
  First we show that all leaf nodes of $T_{\Delta}$ are in orbit
  $\Delta$. Any node $\alpha$ of $T_{\Delta}$ must lie on a path such
  that the endpoints are in orbit $\Delta$. If $\alpha$ is a leaf of
  $T_{\Delta}$, this can only happen when $\alpha$ itself is in
  $\Delta$.

  We will prove the converse by contradiction. If possible let
  $\alpha$ be a vertex in the orbit $\Delta$ which is a not a leaf of
  $T_{\Delta}$. Vertex $\alpha$ must lie on the path between two
  leaves $\beta$ and $\gamma$. Also since $\beta$ and $\gamma$ are
  leaves of $T_\Delta$, they are in the orbit $\Delta$.

  Let $g$ be an automorphism of $T$ which maps $\alpha$ to $\beta$.
  Such an automorphism exists because $\alpha$ and $\beta$ are in the
  same orbit $\Delta$. The image $\alpha^g=\beta$ must lie on the path
  between $\beta^g$ and $\gamma^g$ and neither $\beta^g$ or $\gamma^g$
  is $\beta$. This is impossible because $\beta$ is a leaf of
  $T_\Delta$. 
\end{proof}

\begin{lemma}\label{lem-subtree}
  Let $\gamma$ be a vertex in orbit $\Sigma$. If $\gamma$ is a vertex
  of the subtree $T_{\Delta}$ then subtree $T_{\Sigma}$ is a subtree
  of $T_{\Delta}$.
\end{lemma}
\begin{proof}
  Assume that $\Delta$ is different from $\Sigma$, for otherwise the
  proof is trivial. First we show that all the vertices of $\Sigma$
  are vertices of $T_\Delta$. The vertex $\gamma$ lies on a path
  between two vertices of $\Delta$, say $\alpha$ and $\beta$. Take any
  vertex $\gamma'$ from the orbit $\Sigma$. There is an automorphism
  $g$ of $T$ which maps $\gamma$ to $\gamma'$. Now $\gamma'=\gamma^g$
  lies on the path between $\alpha^g$ and $\beta^g$ and hence is in
  the tree $T_\Delta$.

  Consider any edge $e$ of $T_\Sigma$. There exists $\gamma_1$ and
  $\gamma_2$ of $\Sigma$ such that $e$ is on the path from $\gamma_1$
  to $\gamma_2$. By previous argument, $T$ contains $\gamma_1$ and
  $\gamma_2$. Since $T$ is a tree, this path is unique and any
  subgraph of $T$, in which $\gamma_1$ and $\gamma_2$ are connected,
  must contain this path. Hence $T_{\Delta}$ contains $e$.
\end{proof}

\begin{lemma}\label{lem-edge}
  Let $T$ be any tree then either there exists a vertex $\alpha$ that
  is fixed by all the automorphisms of $T$ or there exists two
  vertices $\alpha$ and $\beta$ connected via an edge $e$ such that
  $\{ \alpha,\beta\}$ is an orbit of $\Aut{T}$. In the latter case
  every automorphism maps $e$ to itself.
\end{lemma}



\begin{proof}
  Consider the following partial order between orbits of $\Aut{T}$:
  $\Sigma \leq \Delta$ if $T_\Sigma$ is a subtree of $T_\Delta$. The
  relation $\leq$ is clearly a partial order because the ``subtree''
  relation is.  Since there are finitely many orbits there is always a
  minimal orbit under the above ordering.  From Lemmas~\ref{lemleaf}
  and \ref{lem-subtree} it follows that for an orbit $\Delta$, if
  $\Sigma$ is the orbit containing an internal node $\gamma$ of
  $T_\Delta$ then $\Sigma$ is strictly less than $\Delta$. Therefore
  for any minimal orbit $\Delta$, all the nodes are leaves. This is
  possible if either $T_\Delta$ a singleton vertex $\alpha$ or
  consists of exactly two nodes connected via an edge. In the former
  case all automorphisms of $T$ have to fix $\alpha$, whereas in the
  latter case the two nodes may be flipped but the edge connecting
  them has to be mapped to itself.
\end{proof}

It follows from Lemma~\ref{lem-edge} that any tree $T$ can be rooted,
either at a vertex or at an edge with out changing the automorphism.
Given a tree $T$, since computing the a generating set for $\Aut{T}$
can be done in polynomial time, we can determine all the orbits of
$\Aut{T}$ by a simple transitive closure algorithm. Having computed
these orbits, we determine whether $T$ has singleton orbit or an orbit
of cardinality $2$. For trees with an orbit containing a single vertex
$\alpha$, rooting the tree at $\alpha$ does not change the
automorphism group. On the other hand if the tree has an orbit with
two elements we can add a dummy root as in
Figure~\ref{fig-maximal-orbit} without changing the automorphism
group. Since by Lemma~\ref{lem-edge} these are the only two
possibilities we have the following theorem.

\begin{theorem}
  There is a polynomial time algorithm that, given as input a tree $T$,
  outputs a rooted tree $T'$ such that for any group $G$, $G$ is
  representable on $T$ if and only if $G$ is representable on the rooted tree
  $T'$.
\end{theorem}

For the rest of the section by a tree we mean a rooted tree. We will
prove the reduction from representability on rooted trees to
permutation representability. First we characterise the automorphism
group of a tree in terms of wreath product [Theorem \ref{thmtreeauto}]
and then show that we can find a nontrivial homomorphism, if there
exists one, from the given group $G$ to this automorphism group by
querying a permutation representability oracle.

\begin{definition}[Semidirect product and wreath product]
  Let $G$ and $A$ be any two group and let $\varphi$ be any
  homomorphism from $G$ to $\Aut{A}$, then the semi-direct product
  $G\ltimes_\varphi A$ is the group whose underlying set is $G\times
  H$ and the multiplication is defined as $(g,a) (h,b) = (gh,
  a^{\varphi(h)}b)$.
  
  We use $W_n(A)$ to denote the wreath product $S_n \wr A$ which is
  the semidirect product $S_n\ltimes_\varphi A^n$, where $A^n$ is the
  $n$-fold direct product of $A$ and ${\varphi(h)}$, for each $h$ in
  $S_n$, permutes $\mathbf{a} \in A^n$ according to the permutation
  $h$, i.e. maps $(\ldots,a_i,\ldots) \in A^n$ to
  $(\ldots,a_{j},\ldots)$ where $j^h = i$.
\end{definition}

As the wreath product is a semidirect product, we have the following
lemma.

\begin{lemma}\label{lem-wreath-property}
  The wreath product $W_n(A)$ contains (isomorphic copies of) $S_n$
  and $A^n$ as subgroups such that $A^n$ is normal and the quotient
  group $W_n(A)/A^n=S_n$.
\end{lemma}

For the rest of the section fix the following: Let $T$ be a tree with
root $\omega$ with $k$ children. Consider the subtrees of $T$ rooted
at each of these $k$ children and partition them such that two
subtrees are in the same partition if and only if they are
isomorphic. Let $t$ be the number of partitions and let $k_i$, for $(1
\leq i \leq t)$, be the number of subtrees in the $i$-th
partition. For each $i$, pick a representative subtree $T_i$ from the
$i$-th partition and let $A_i$ denote the automorphism group of
$T_i$. The following result is well known but a proof is given for
completeness.

\begin{theorem}\label{thmtreeauto}
  The automorphism group of the tree $T$ is (isomorphic to) the direct
  product $\prod_{i=1}^t W_{k_i}(A_i)$.
\end{theorem}
\begin{proof}
  Let $\omega_1,\ldots,\omega_k$ be the children of the root $\omega$
  and let $X_i$ denote the subtree rooted at $\omega_i$. We first
  consider the case when $t=1$, i.e. all the subtrees $X_i$ are
  isomorphic. Any automorphism $g$ of $T$ must permute the children
  $\omega_i$'s among themselves and whenever $\omega_i^g = \omega_j$,
  the entire subtree $X_i$ maps to $X_j$.  As all the subtrees $X_i$
  are isomorphic to $T_1$, the forest $\{X_1,\ldots, X_k\}$ can be
  thought of as the disjoint union of $k$ copies of the tree $T_1$ by
  fixing, for each $i$, an isomorphism $\sigma_i$ from $T_1$ to $X_i$.
  
  For an automorphism $g$ of $T$, define the permutation $\tilde{g}\in
  S_k$ and the automorphisms $a_i(g)$ of $T_1$ as follows: if
  $\omega_i^g = \omega_j$ then $i^{\tilde{g}} = j$ and $a_i(g) =
  \sigma_i g \sigma_j^{-1}$.  Consider the map $\phi$ from $\Aut{T}$
  to $W_k(A)$ which maps an automorphism $g$ to the group element
  $(\tilde{g},a_1(g),\ldots,a_k(g))$ in $W_k (A)$.  It is easy to
  verify that $\phi$ is the desired isomorphism.

  When the number of partitions $t$ is greater than $1$, any
  automorphism of $T$ fixes the root $\omega$ and permutes the
  subtrees in the $i$-th partition among themselves. Therefore the
  automorphism group of $T$ is same as the automorphism group of the
  collection of forests $F_i$ one for each partition $i$. Each forest
  is a disjoint union of $k_i$ copies of $T_i$ and we can argue as
  before that its automorphism group is (isomorphic to)
  $W_{k_i}(A)$. Therefore $\Aut{T}$ should be the direct product
  $\prod_{i=1}^t W_{k_i}(A_i)$.
\end{proof}

\begin{lemma}\label{lem-break1}
  If the group $G$ can be represented on the tree $T$, then there
  exists $1\leq i \leq t$ such that there is a nontrivial homomorphism
  from $G$ to $W_{k_i}(A_i)$.
\end{lemma}

\begin{proof}
  If there is a nontrivial homomorphism from a group $G$ to the direct
  product of groups $H_1,\ldots,H_t$ then for some $i$, $1 \leq i \leq
  t$, there is a nontrivial homomorphism from $G$ to $H_i$. The lemma
  then follows from Theorem \ref{thmtreeauto}.
\end{proof}

\begin{lemma}\label{lem-break2}
  If there is a nontrivial homomorphism $\rho$ from a group $G$ to
  $W_n(A)$ then there is also a nontrivial homomorphism from $G$
  either to $S_n$ or to $A$.
\end{lemma}
\begin{proof}
  Let $\rho$ be a nontrivial homomorphism $G$ to $W_n(A)$. Since $A^n$
  is a normal subgroup of $W_n(A)$ and the quotient group $W_n(A)/A^n$
  is $S_n$, there is a homomorphism $\rho'$ from $W_n(A)$ to $S_n$
  with kernel $A^n$.  The composition of $\rho$ and $\rho'$ is a
  homomorphism from $G$ to $S_n$.

  If $\rho'\cdot\rho$ is trivial then $\rho'$ maps all elements of
  $\rho(G)$ to identity of $S_n$. Which imply that $\rho(G)$ is a
  subgroup of the kernel of $\rho'$, that is $A^n$. So, $\rho$ is a
  nontrivial homomorphism from $G$ to $A^n$. Hence there must be a
  nontrivial homomorphism from $G$ to $A$.
\end{proof}

\begin{theorem}
  Given a group $G$ and a rooted tree $T$ with $n$ nodes and an oracle
  for deciding whether $G$ has a nontrivial homomorphism to $S_m$ for
  $1 \leq m \leq n$, it can be decided in polynomial time whether $G$
  can be represented on $T$.
\end{theorem}

\begin{proof}
  If the tree has only one vertex then reject. Otherwise let $t$,
  $k_1,\ldots, k_t$ and $A_1,\ldots A_t$ be the quantities as defined
  in Theorem~\ref{thmtreeauto}. Since there is efficient algorithm to
  compute tree isomorphism, $t$ and $k_1,\ldots, k_t$ can be computed
  in polynomial time. If $G$ is representable on $T$ then, by
  Lemma~\ref{lem-break1} and Lemma~\ref{lem-break2}, there is a
  nontrivial homomorphism form $G$ to either $S_{k_i}$ or $A_i$ for
  some $i$. Using the oracle, check whether there is a nontrivial
  homomorphism to any of the symmetric groups. If found then accept,
  otherwise for all $i$, decide whether there is a nontrivial
  homomorphism to $A_i$ by choosing a subtree $T_i$ from the $i^{th}$
  partition and recursively asking whether $G$ is representable on
  $T_i$. Total number of recursive calls is bounded by the number of
  vertices of $T$. Hence the reduction is polynomial time.
\end{proof}

\section{Conclusion}

In this paper we studied the group representability problem, a
computational problem that is closely related to graph
isomorphism. The representability problem could be equivalent to graph
isomorphism, but the results of
Section~\ref{sect-tree-representability} give some, albeit weak,
evidence that this might not be the case. It would be interesting to
know what is the exact complexity of this problem vis a vis the graph
isomorphism problem. We know from the work of
Mathon~\cite{mathon79note} that the graph isomorphism problem is
equivalent to its functional version where, given two graphs $X$ and
$Y$, we have to compute an isomorphism if there exists one. The
functional version of group representability, namely give a group $G$
and a graph $X$ compute a nontrivial representation if it exists, does
not appear to be equivalent to the decision version. Also it would be
interesting to know if the representability problem shares some of
lowness of graph
isomorphism~\cite{schoning87graph,kobler92graph,arvind2002graph}.  Our
hope is that, like the study of group representation in geometry and
mathematics, the study of group representability on graphs help us
better understand the graph isomorphism.

\bibliographystyle{plain}
\bibliography{./bibdata}

\end{document}